\documentclass[conference]{IEEEtran}
\IEEEoverridecommandlockouts
\usepackage{cite}
\usepackage{amsmath,amssymb,amsfonts}
\usepackage{algorithmic}
\usepackage{graphicx}
\usepackage{textcomp}
\usepackage{subcaption}
\usepackage{svg}
\usepackage{quantikz}
\def\BibTeX{{\rm B\kern-.05em{\sc i\kern-.025em b}\kern-.08em
    T\kern-.1667em\lower.7ex\hbox{E}\kern-.125emX}}

\usepackage{authblk}
\DeclareMathOperator*{\argmin}{arg\,min}

\begin{document}


\title{Two-Qubit Implementation of QAOA for MAX-CUT on an NV-Center Quantum Processor}
\author[1, 3]{Leon E. Röscher}
\author[1, 3]{Talía L. M. Lezama}
\author[3, 4]{Luca Cimino}
\author[1, 3]{Jonah vom Hofe}
\author[1,2,3]{\authorcr Riccardo Bassoli} 
\author[1,2]{Frank H. P. Fitzek}

\affil[1]{Deutsche Telekom Chair of Communication Networks, Technische Universität Dresden, Germany}
\affil[2]{Centre for Tactile Internet with Human-in-the-Loop (CeTI), Dresden, Germany}
\affil[3]{Quantum Communication Networks Research Group, Technische Universität Dresden, Germany}
\affil[4]{Department of Computer Science and Engineering, University of Bologna, Italy}

\maketitle

\begin{abstract}
We report a proof-of-principle implementation of the quantum approximate
optimization algorithm (QAOA) for the smallest nontrivial MAX-CUT instance on an
NV-center-based quantum processor operating at room temperature. The two-qubit
register is encoded in the electron spin and the \({}^{14}\mathrm{N}\) nuclear
spin of a single NV\(^-\) center. Using a minimization formulation of MAX-CUT,
we implement a single-layer QAOA ansatz with native entangling and single-qubit
control operations. Because the optical readout of the NV\(^-\) center is not
projective in the computational basis, we reconstruct computational-basis
populations from averaged fluorescence signals and use them to determine the
experimental QAOA cost landscape by scanning the variational parameters. These
results show that the core elements of QAOA can be realized on this platform and
establish a baseline for future improvements in phase tracking, coherence-
preserving control, and scaling to larger problem sizes.
\end{abstract}


\section{Introduction}

Noisy Intermediate-Scale Quantum (NISQ) devices are quantum processors with
limited qubit numbers and non-negligible noise, for which deep fault-tolerant
circuits remain out of reach~\cite{preskill2018quantum}. Within this regime,
variational quantum algorithms have emerged as a promising approach because they
combine shallow parameterized quantum circuits with classical optimization.
Among these, the Quantum Approximate Optimization Algorithm (QAOA), introduced
by Farhi et al.~\cite{farhi2014quantumapproximateoptimizationalgorithm}, is one
of the most widely studied methods for combinatorial optimization on near-term
hardware.

In this work, we focus on MAX-CUT, a standard NP-hard combinatorial problem that
serves as a natural benchmark for QAOA. Its graph-based structure makes it
particularly suitable for studying the relationship between a problem
Hamiltonian and the connectivity of a quantum processing unit (QPU). At the same
time, present-day hardware constraints, including limited qubit number,
restricted connectivity, and control noise, make even small proof-of-principle
implementations valuable for understanding the practical feasibility of QAOA on
specific platforms.

Nitrogen-vacancy (NV) centers in diamond are a promising solid-state platform
for quantum information processing because they can operate at room temperature
and support coherent control of coupled electron and nuclear
spins~\cite{balasubramanian_nanoscale_2008}. Early demonstrations on NV-center
systems include the room-temperature implementation of the Deutsch--Jozsa
algorithm using a single electronic spin in diamond
\cite{PhysRevLett.105.040504}. Later, a programmable two-qubit solid-state
quantum processor based on a single NV center was demonstrated under ambient
conditions~\cite{wu2019programmable}. These results establish NV centers as a
viable platform for small-scale quantum information processing, while their
optical interface also makes them attractive for modular and networked quantum
architectures \cite{Childress_Hanson_2013,jager2025modeling}.

QAOA and related variational methods have been studied extensively, both
theoretically and experimentally. Beyond the original framework
\cite{farhi2014quantumapproximateoptimizationalgorithm}, generalizations to
constrained optimization and hardware demonstrations on other platforms have
shown that performance depends strongly on the gate set, connectivity, and noise
characteristics of the underlying processor
\cite{hadfield2019quantum,pagano2020quantum}.

Despite this progress, the feasibility of implementing QAOA for combinatorial
optimization on room-temperature NV-center hardware remains largely unexplored.
In this work, we investigate a proof-of-principle implementation of QAOA for the
smallest nontrivial MAX-CUT problem on an NV\(^-\)-center quantum processor. We
map the problem to a hardware-adapted QAOA circuit, implement the corresponding
control sequence experimentally, reconstruct computational-basis populations
from fluorescence readout, and compare the resulting experimental cost landscape
with the corresponding ideal prediction.

\section{QAOA Formulation of MAX-CUT}

The quantum approximate optimization algorithm was introduced as a hybrid
quantum-classical algorithm for approximately solving combinatorial optimization
problems~\cite{farhi2014quantumapproximateoptimizationalgorithm}. Candidate
solutions are encoded as bit strings \(x\in\{0,1\}^n\), for example representing
a graph partition or two-coloring. In this work, we use a minimization
formulation, which is natural in the Hamiltonian language of energy minimization
and ground states used in many-body physics. We therefore consider a cost
function \(C:\{0,1\}^n\to\mathbb{R}\) and define the solution as
\begin{equation}
    x^*=\argmin_{x\in\{0,1\}^n} C(x).
\end{equation}
A brute-force search over all \(2^n\) candidate strings scales exponentially
with \(n\), and no exact polynomial-time algorithm is known for NP-hard problems
such as MAX-CUT.

For a finite simple undirected graph \(G=(V,E)\) with vertex set \(V\), edge set
\(E\), and adjacency matrix \(A\), where \(A_{ij}=A_{ji}\) and \(A_{ii}=0\),
MAX-CUT seeks to find a partitioning of the vertex set into two disjoint subsets
\(S\) and \(T\) that maximizes the number of edges between them. A candidate
partition can be represented by a bit string~\(x=(x_1,\dots,x_{|V|})^T\), where
\(|V|\) is the number of vertices in the graph and
\begin{equation}
x_i=
\begin{cases}
0,& \text{if } i\in S,\\
1,& \text{if } i\in T.
\end{cases}
\label{eq:partition}
\end{equation}
The corresponding cut value is
\begin{equation}
    C_{\mathrm{cut}}(x)=\sum_{i<j}A_{ij}\bigl(x_i+x_j-2x_ix_j\bigr),
    \label{eq:cut_value}
\end{equation}
where each edge is counted exactly once. Since MAX-CUT is naturally a
maximization problem, one seeks a partition that maximizes
\(C_{\mathrm{cut}}(x)\). In the present work, however, we use an equivalent
minimization formulation and define the cost function
\begin{equation}
C(x)=-C_{\mathrm{cut}}(x)
=-\sum_{i<j}A_{ij}\bigl(x_i+x_j-2x_ix_j\bigr).
\label{eq:cost_function}
\end{equation}
Thus, minimizing \(C(x)\) is equivalent to maximizing the cut value. It is often
convenient to map the binary variables to spin variables \(z_i\in\{-1,+1\}\) via
\begin{equation}
    x_i=\frac{z_i+1}{2}.
    \label{eq:map_bit_spin}
\end{equation} 
With this substitution, the cost function becomes
\begin{equation}
   C(z)=-\frac{1}{2}\sum_{i<j} A_{ij}\bigl(1-z_i z_j\bigr). 
   \label{eq:cost_function_spin}
\end{equation}
The corresponding cost Hamiltonian is
\begin{equation}
    H_C=-\frac{1}{2}\sum_{i<j} A_{ij}\bigl(I-Z_iZ_j\bigr),
    \label{eq:cost_Hamiltonian}
\end{equation}
which is diagonal in the computational basis. Here \(I\) denotes the identity
operator and \(Z_i\) denotes the Pauli-\(Z\) operator acting on qubit~\(i\). For
every computational basis state \(|x\rangle\), one has
\begin{equation}
    \langle x|H_C|x\rangle = C(x),
    \label{eq:hc_expectation_cost}
\end{equation}
where \(C(x)\) and \(C(z)\) denote the same cost function in binary and spin
variables, respectively, with \(z_i=2x_i-1\).

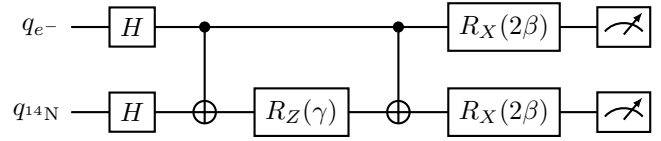
\begin{figure}
\begin{quantikz}[row sep=0.5cm, column sep=0.5cm]
\lstick{$q_{e^-}$} & \gate{H} & \ctrl{1} & & \ctrl{1} & \gate{R_X(2\beta)} & \meter{} \\
\lstick{$q_{{}^{14}\mathrm{N}}$} & \gate{H} & \targ{} & \gate{R_Z(\gamma)} & \targ{} & \gate{R_X(2\beta)} & \meter{}
\end{quantikz}
\caption{QAOA circuit for the two-vertex graph with \(p=1\), implemented on
the electron spin (\(q_{e^-}\)) and nitrogen nuclear spin
(\(q_{{}^{14}\mathrm{N}}\)) of an NV\(^-\) center. The
CNOT-\(R_Z(\gamma)\)-CNOT block realizes \(RZZ(\gamma)\), and the final
\(R_X(2\beta)\) gates on each qubit implement the mixing unitary.}
\label{fig:qaoa_circuit}
\end{figure}

The associated QAOA cost unitary is
\begin{equation}
    U_C(\gamma)=e^{-i\gamma H_C}.
    \label{eq:cost_unitary}
\end{equation}
Using the convention \(RZZ(\theta)=e^{-i\theta Z\otimes Z/2}\), each nontrivial
factor can be written as \(\exp\!\left(\frac{i\gamma}{2}(I-Z_iZ_j)\right) =
e^{i\gamma/2}RZZ_{i,j}(\gamma)\), and noting that only terms with \(A_{ij}=1\)
contribute, one obtains
\begin{equation}
    U_C(\gamma)\doteq \prod_{i<j\,:\,A_{ij}=1} RZZ_{i,j}(\gamma),
    \label{eq:cost_unitary_gamma}
\end{equation}
where \(\doteq\) denotes equality up to a global phase. Equivalently, this
corresponds to one \(RZZ\) gate for each edge of the graph.

The mixing unitary is
\begin{equation}
    U_B(\beta)=e^{-i\beta\sum_{i=1}^{|V|}X_i}
          = \bigl(R_X(2\beta)\bigr)^{\otimes |V|},
    \label{eq:mixing_unitary}
\end{equation}
where \(X_i\) denotes the Pauli-\(X\) operator acting on qubit~\(i\).

The QAOA ansatz is initialized in the uniform superposition state
\(|+\rangle^{\otimes |V|}\), which can equivalently be prepared as
\begin{equation}
   |+\rangle^{\otimes |V|}=H^{\otimes |V|}|0\cdots 0\rangle.
   \label{eq:initial_sup_state}
\end{equation}
For variational parameters \( \boldsymbol{\beta}=(\beta_1,\dots,\beta_p),
\boldsymbol{\gamma}=(\gamma_1,\dots,\gamma_p), \) the depth-\(p\) ansatz is
given by
\begin{equation}
|\psi(\boldsymbol{\beta},\boldsymbol{\gamma})\rangle
=
U_B(\beta_p)U_C(\gamma_p)\cdots U_B(\beta_1)U_C(\gamma_1)\,|+\rangle^{\otimes |V|}.
\label{eq:ansatz}
\end{equation}

QAOA proceeds by preparing and sampling this ansatz. A measurement in the
computational basis, equivalently in the joint \(Z\)-basis, yields a classical
bit string~\(x\in\{0,1\}^{|V|}\), which specifies a partition of the graph.
Repeated measurements estimate the probability distribution
\(p_{\boldsymbol{\beta},\boldsymbol{\gamma}}(x)\) over candidate solutions. For
fixed parameters \((\boldsymbol{\beta},\boldsymbol{\gamma})\), the expected cost
is
\begin{equation}
    F(\boldsymbol{\beta},\boldsymbol{\gamma})
=
\langle \psi(\boldsymbol{\beta},\boldsymbol{\gamma})|H_C|\psi(\boldsymbol{\beta},\boldsymbol{\gamma})\rangle=\sum_{x\in\{0,1\}^{|V|}} C(x)\,p_{\boldsymbol{\beta},\boldsymbol{\gamma}}(x).
\label{eq:expected_cost}
\end{equation}
A classical optimizer is then used to minimize
\(F(\boldsymbol{\beta},\boldsymbol{\gamma})\).

\begin{figure*}[t]
\centering

\begin{subfigure}{\textwidth}
    \centering
        \phantomcaption
    \makebox[0pt][l]{\hspace{-0.0em}\raisebox{9.0em}{\textbf{(a)}}}%
    \includegraphics[width=\textwidth]{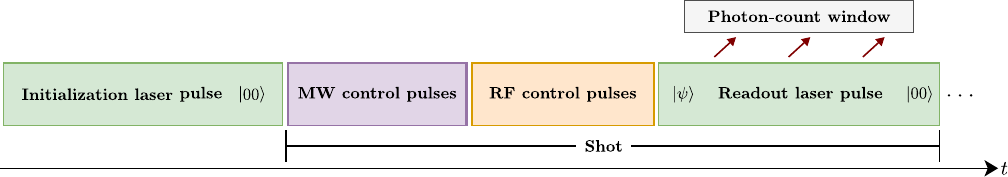}
    \label{fig:measurement_cycle}
\end{subfigure}

\vspace{0.2cm}

\begin{subfigure}{\textwidth}
    \centering
    \phantomcaption
    \makebox[0pt][l]{\hspace{-0.4em}\raisebox{3.6em}{\textbf{(b)}}}%
    \scalebox{0.85}{%
    \begin{tabular}{cccc}
        \begin{quantikz}[row sep=0.25cm, column sep=0.2cm]
            \lstick{$|\phi_{1}\rangle$} & \qw      & \gate[2,style={dashed,rounded corners,fill=gray!10}]{\shortstack{optical\\readout}}\arrow[r]
                & \rstick{$\langle O\rangle_{|\phi\rangle}$} \\
            \lstick{$|\phi_{2}\rangle$} & \qw      & \qw
        \end{quantikz}
        &
        \begin{quantikz}[row sep=0.25cm, column sep=0.2cm]
            \lstick{$|\phi_{1}\rangle$} & \gate{X} & \gate[2,style={dashed,rounded corners,fill=gray!10}]{\shortstack{optical\\readout}} \arrow[r]
                & \rstick{$\langle O\rangle_{X_1|\phi\rangle}$} \\
            \lstick{$|\phi_{2}\rangle$} & \qw      & \qw
        \end{quantikz}
        &
        \begin{quantikz}[row sep=0.25cm, column sep=0.2cm]
            \lstick{$|\phi_{1}\rangle$} & \qw      & \gate[2,style={dashed,rounded corners,fill=gray!10}]{\shortstack{optical\\readout}} \arrow[r]
                & \rstick{$\langle O\rangle_{X_2|\phi\rangle}$} \\
            \lstick{$|\phi_{2}\rangle$} & \gate{X} & \qw
        \end{quantikz}
        &
        \begin{quantikz}[row sep=0.25cm, column sep=0.2cm]
            \lstick{$|\phi_{1}\rangle$} & \gate{X} & \gate[2,style={dashed,rounded corners,fill=gray!10}]{\shortstack{optical\\readout}}\arrow[r]
                & \rstick{$\langle O\rangle_{X_1X_2|\phi\rangle}$} \\
            \lstick{$|\phi_{2}\rangle$} & \gate{X} & \qw
        \end{quantikz}
    \end{tabular}}
\label{fig:population_reconstruction_circuits}
\end{subfigure}
\caption{\textbf{Overview of the measurement protocol.}
\textbf{(a)}~One measurement shot consists of optical initialization, coherent
control, and optical readout. Microwave pulses address the electron spin,
conditional radio-frequency pulses address the nuclear spin, and the detected
fluorescence is recorded as a running average of the photon count per shot.
\textbf{(b)}~Readout circuits used for calibration and population
reconstruction. The final operations \(I\), \(X_1\), \(X_2\), and \(X_1X_2\)
generate four mean photon counts \(\bar n\), which are used to determine the
calibration values \(\{I_s\}\) and to reconstruct the computational-basis
populations according to Eq.~\eqref{eq:four_measurements}. The circuits are
executed in alternating order to reduce drift.
}
\end{figure*}
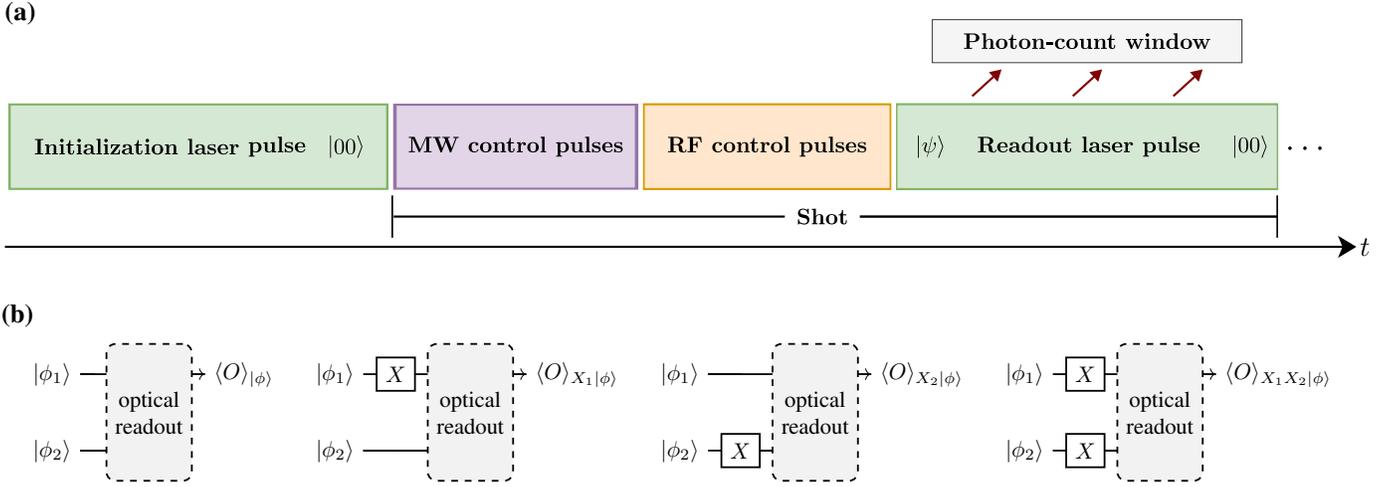

\section{Two-Qubit Implementation on the NV-Center Processor}

The fundamental two-qubit entangling operation required to implement MAX-CUT
QAOA is the \(RZZ\) gate, as follows from Eq.~\eqref{eq:cost_unitary_gamma}. To
demonstrate the feasibility of this approach on an NV\(^-\)-center-based quantum
processor, we consider the simplest nontrivial instance, namely the two-vertex
graph
\[
G=\bigl(\{e^-,\,{}^{14}\mathrm{N}\},\;\{\{e^-,\,{}^{14}\mathrm{N}\}\}\bigr),
\]
with a single QAOA layer (\(p=1\)). This instance is realized using the electron
spin \(e^-\) and the \({}^{14}\mathrm N\) nuclear spin of a single NV\(^-\)
center.

Restricting the general depth-\(p\) ansatz of Eq.~\eqref{eq:ansatz} to this
graph and to \(p=1\) gives
\begin{equation}
    |\psi(\beta,\gamma)\rangle
    =
    U_B(\beta)\,U_C(\gamma)\,H^{\otimes 2}|00\rangle,
    \label{eq:two_qubit_ansatz}
\end{equation}
where \(|00\rangle\) denotes the initialized computational basis state defined
in Eq.~\eqref{eq:computational_basis_nv} of Sec.~\ref{sec:physical_setup}. In
this case, the cost unitary reduces to a single two-qubit interaction,
\begin{equation}
U_C(\gamma)=RZZ_{e^-,\,{}^{14}\mathrm N}(\gamma),    
\end{equation}
acting on the two spins. The \(RZZ\) gate is decomposed into native operations
as \( RZZ(\gamma) = \mathrm{CNOT}\cdot\bigl(I\otimes
R_Z(\gamma)\bigr)\cdot\mathrm{CNOT}\), while the mixing unitary acts as
\begin{equation}
    U_B(\beta)=R_X(2\beta)^{\otimes 2}.
\end{equation}
The resulting circuit is shown in Fig.~\ref{fig:qaoa_circuit}, where the angles
\(\gamma\) and \(\beta\) correspond directly to the QAOA variational parameters.

\subsection{Description of the physical setup}\label{sec:physical_setup}

The quantum processor used in this work is based on a single NV\(^-\) center in
diamond and is operated at room temperature. Control is implemented through
microwave, radio-frequency, and optical pulses. The NV\(^-\) center is optically
pumped by green laser pulses, and the emitted red fluorescence is detected by an
avalanche photodiode (APD)~\cite{balasubramanian_nanoscale_2008}.

Physically, the electron and the \({}^{14}\mathrm N\) nuclear spin are both
spin-1 systems, with spin quantum numbers \(S=1\) and \(I=1\), respectively. In
the present work, these degrees of freedom are restricted to the two-level
subspaces \(\{|m_s=0\rangle, |m_s=-1\rangle\}\) and \(\{|m_I=+1\rangle,
|m_I=0\rangle\}\), respectively, which together define the effective two-qubit
register.

The electron-spin transition \(|m_s=0\rangle\leftrightarrow|m_s=-1\rangle\),
located at approximately \(1.47\,\mathrm{GHz}\) under the applied magnetic
field, is driven by microwave pulses. The pulse amplitude is calibrated from
Rabi oscillations, yielding a \(\pi\)-pulse duration of \(t_\pi\approx
104\,\mathrm{ns}\). The rotation axis in the equatorial plane of the Bloch
sphere is determined by the phase of the microwave drive relative to the
rotating frame, while the rotation angle is set by the pulse amplitude and
duration. In our convention, a drive phase of \(0\) implements rotations about
the \(x\)-axis, while a phase of \(\pi/2\) implements rotations about the
\(y\)-axis. For example, a calibrated \(\pi\)-pulse with phase \(0\) realizes
\(R_x(\pi)\), i.e.\ an \(X\) gate. Rotations about the \(z\)-axis are
implemented virtually by updating the phase of the rotating reference frame;
since no physical pulse is applied, these operations incur no additional pulse
time or decoherence.

The device is operated near the excited-state level anticrossing (ESLAC), at an
applied magnetic field of approximately \(B\sim 500\,\mathrm{G}\). In this
regime, hyperfine-mediated electron-nuclear flip-flop processes enable optical
transfer of polarization from the electron spin to the \({}^{14}\mathrm N\)
nuclear spin, thereby preferentially preparing the state
\(|m_s=0,\,m_I=+1\rangle\)~\cite{he2024direct}. In the present work, this
operating point was inherited from the hardware configuration, while the
experimental procedures focused on calibration, qubit characterization, and gate
measurements.

Within the computational subspace \(\{|m_I=+1\rangle, |m_I=0\rangle\}\), the
relevant conditioned drives address the transition
\(|m_I=0\rangle\leftrightarrow|m_I=+1\rangle\), whose frequency depends on the
electron-spin state. In our setup, these two conditional drives are calibrated
at approximately \(5.1\,\mathrm{MHz}\) for \(m_s=0\) and \(2.9\,\mathrm{MHz}\)
for \(m_s=-1\). Both transitions have \(\pi\)-pulse durations on the order of
\(t_\pi\approx 13\,\mu\mathrm{s}\). Unconditioned nuclear-spin \(R_x\) rotations
are realized by driving both transitions simultaneously, thereby rendering the
operation effectively independent of the electron-spin state. Nuclear-spin
\(R_y\) and \(R_z\) rotations are implemented through phase control of the
radio-frequency drives, analogously to the electron-spin case. These two
conditioned drives play the role of the native controlled operations used in the
pulse-level implementation.

We identify the experimentally prepared state \(|m_s=0,\,m_I=+1\rangle\) with
the computational basis state \(|00\rangle\). Under this convention, the
effective two-qubit basis induced by the chosen subspaces is
\begin{equation}
\begin{aligned}
&|m_s=0,\,m_I=+1\rangle \equiv |00\rangle,\quad |m_s=0,\,m_I=0\rangle \equiv |01\rangle,\\
&|m_s=-1,\,m_I=+1\rangle \equiv |10\rangle, \,|m_s=-1,\,m_I=0\rangle \equiv |11\rangle.
\end{aligned}
\label{eq:computational_basis_nv}
\end{equation}
Accordingly, the initial Hadamard layer in Eq.~\eqref{eq:two_qubit_ansatz}
prepares the uniform superposition \(H^{\otimes 2}|00\rangle\) from this
experimentally defined computational input state.

\begin{figure}[t!]
    \centering
    \includegraphics[width=\columnwidth]{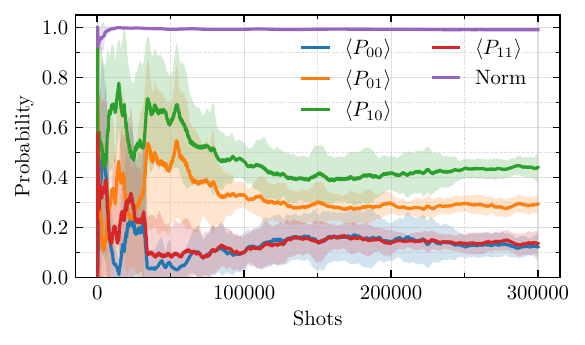}
    \caption{Convergence of the reconstructed computational-basis populations
    for the ansatz \(|\psi(0.15\pi,1.5\pi)\rangle\) as a function of the
    accumulated number of shots. Solid lines denote the mean reconstructed
    populations, while the shaded regions show the corresponding standard
    deviation. The norm, defined as the sum of the four reconstructed
    populations (before avering), is shown analogously.}
    \label{fig:populations_convergence}
\end{figure}

\begin{figure*}[h!]
\centering
\includegraphics[width=\textwidth]{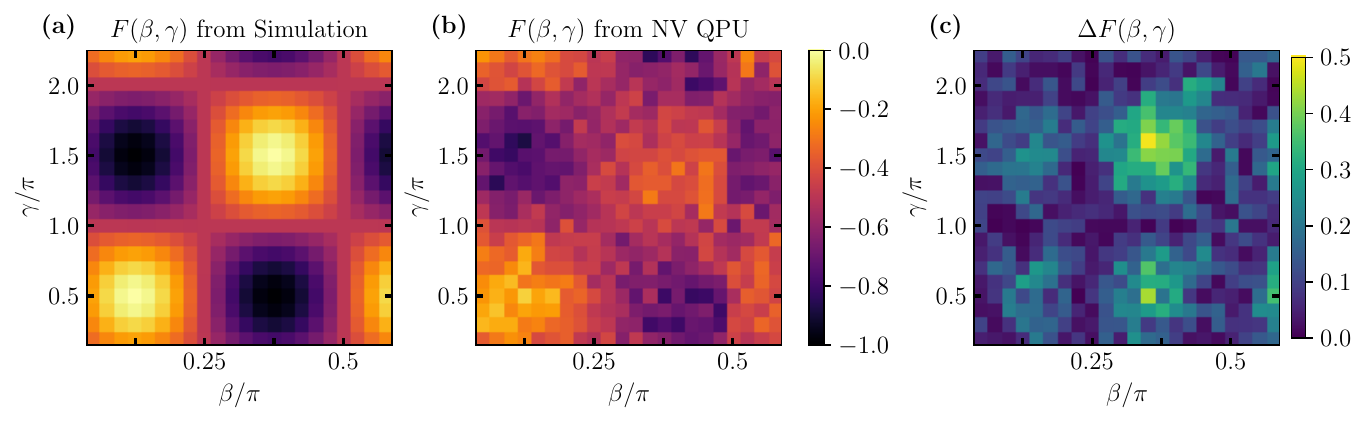}
\caption{\textbf{QAOA cost landscapes for the two-vertex MAX-CUT instance}.
\textbf{(a)}~The ideal QAOA cost landscape obtained from a noise-free
state-vector simulation, \textbf{(b)}~the experimentally reconstructed
landscape measured on the NV-center quantum processor, and
\textbf{(c)}~the absolute difference between the two.}
\label{fig:cost_landscape}
\end{figure*}

\subsection{Measurement and population reconstruction}\label{sec:measurement}

The optical readout of the NV\(^-\) center does not yield single-shot projective
measurements in the computational basis. In this work, we use the term
\emph{shot} to denote one repetition of the pulse sequence followed by optical
readout. Each shot yields an integer photon count, but the overlap of the
photon-count distributions for different states prevents reliable single-shot
state discrimination. The relevant quantity is therefore the mean photon count
per shot,
\begin{equation}
    \bar{n} = \frac{1}{N}\sum_{k=1}^{N} n_k,
    \label{eq:mean_photon_count_per_shot}
\end{equation}
where \(n_k\) is the photon count recorded in the \(k\)-th shot and \(N\) is the
total number of shots. As \(N\) increases, \(\bar{n}\) converges to a stable
value that depends on the quantum state and can therefore be calibrated for each
basis state. We refer to this quantity as the fluorescence intensity, which is
recorded experimentally as a running average over repeated shots.

To make this precise, we describe the two-qubit readout by the Hermitian
observable
\begin{equation}
    O = \sum_{s\in\{0,1\}^2} I_s\,|s\rangle\langle s|
      = \sum_{s\in\{0,1\}^2} I_s\,P_s,
    \label{eq:measurement_observable}
\end{equation}
where \(P_s = |s\rangle\langle s|\) is the projector onto basis state
\(|s\rangle\), and \(I_s \in \mathbb{R}^+\) are calibration coefficients
determined below. This description identifies the mean fluorescence intensity
associated with a state \(|\psi\rangle\) with the expectation value of \(O\),
\begin{equation}
    \langle O\rangle_{|\psi\rangle}
    = \sum_{s\in\{0,1\}^2} I_s\,\langle P_s\rangle_{|\psi\rangle},
    \label{eq:observable_expectation}
\end{equation}
where \(\langle P_s\rangle_{|\psi\rangle}\) denotes the computational-basis
population of \(|s\rangle\) in the state \(|\psi\rangle\). The calibration
coefficients \(\{I_s\}\) are determined empirically by preparing each basis
state \(|s\rangle\) directly and recording the corresponding mean fluorescence
intensity. The observable \(O\) is then constructed from these measured values,
so that by construction
\begin{equation}
    \langle O\rangle_{|s\rangle} = I_s
    \qquad\text{for all } s\in\{0,1\}^2.
    \label{eq:calibration_values}
\end{equation}

To reconstruct the computational-basis populations for an arbitrary implemented
circuit \(U\), we measure the four expectation values
\begin{equation}
    \langle O\rangle_{U|00\rangle},\quad
    \langle O\rangle_{X_1 U|00\rangle},\quad
    \langle O\rangle_{X_2 U|00\rangle},\quad
    \langle O\rangle_{X_1 X_2 U|00\rangle},
    \label{eq:four_measurements}
\end{equation}
where \(X_k\) denotes a Pauli-\(X\) operation applied to qubit \(k\) after the
circuit \(U\), immediately before readout. The populations can therefore be
recovered by linear inversion, which takes the form of a Walsh-Hadamard
transform~\cite{cooper2016coherent, nielsen2010quantum}; the explicit
reconstruction formula is derived in
Appendix~\ref{app:population_reconstruction}. Since this procedure recovers only
the diagonal elements of the density matrix in the computational basis, we refer
to it as \emph{population reconstruction} rather than full quantum state
tomography.

Using the reconstructed populations, the expected QAOA cost is evaluated as
\begin{equation}
    F(\beta,\gamma)
    = \sum_{x\in\{0,1\}^2} C(x)\,\langle P_x\rangle_{|\psi(\beta,\gamma)\rangle},
    \label{eq:expected_cost_two_qubit}
\end{equation}
which is Eq.~\eqref{eq:expected_cost} restricted to the present two-qubit
implementation. Minimizing \(F(\beta,\gamma)\) over the variational
parameters~\((\beta,\gamma)\) then yields an approximate solution to the
considered MAX-CUT problem.

\section{Experimental Setup and Cost Landscape Characterization}

The NV\(^-\)-center quantum processor used in this work is manufactured by
XeedQ~\cite{xeedq} and controlled with an OPX+ pulse level
controller~\cite{qmach_opx}. The system operates at room temperature with active
stabilization at \(27.5 \pm 0.1\,^\circ\mathrm{C}\). For each measurement batch,
we used the most recent hourly calibration of resonance frequencies and pulse
amplitudes, while keeping the pulse durations fixed in order to preserve a
reproducible timing structure across the scan.

\subsection{Cost landscape characterization}
\label{sec:cost_landscape_characterization}

To map the experimental QAOA cost landscape for the two-vertex MAX-CUT problem,
we performed a grid search over the single-layer (\(p=1\)) ansatz for \( \gamma
\in [\varepsilon\pi,\,(2+\varepsilon)\pi] \) with step size \(0.05\pi\) and \(
\beta \in [\varepsilon\pi,\,(0.5+\varepsilon)\pi] \) with step size
\(0.025\pi\), where \(\varepsilon=0.1\). The offset avoids sampling exactly at
the boundary values, and the chosen ranges cover one full period of the ideal
cost landscape in both parameters.

For each parameter pair \((\beta,\gamma)\), we executed the QAOA circuit in
Fig.~\ref{fig:qaoa_circuit} together with the four calibration circuits and four
reconstruction circuits in Fig.~\ref{fig:population_reconstruction_circuits}.
Each sub-circuit was measured for \(3\times10^{5}\) shots, and four full
realizations per parameter pair, each of which took \(\sim108\)s to compute.

During the sequence, especially during the conditional nuclear-spin gates
(\(\sim 13\,\mu\mathrm{s}\)), the electron spin may accumulate phase relative to
the microwave rotating frame. We do not track this phase explicitly, but assume
that the fixed timing structure makes the resulting offsets reproducible across
the \((\beta,\gamma)\) scan.

From the running-average fluorescence signals, we reconstructed the
computational-basis populations by the Hadamard-based linear-inversion procedure
described in Appendix~\ref{app:population_reconstruction}.
Fig.~\ref{fig:populations_convergence} shows a representative example for
\((\beta,\gamma)=(0.15\pi,1.5\pi)\). The fluctuations decrease substantially
after about \(10^5\) shots, and the sum of the reconstructed populations remains
close to one, reaching \(0.989\pm0.01\) near the final shot count. We therefore
use the final values at \(N=3\times10^5\) as the reconstructed populations for
each parameter pair.

For the two-vertex graph considered here, the expected cost is
\begin{equation}
    F(\beta,\gamma)
    =
    -\bigl(
    \langle P_{01}\rangle_{|\psi(\beta,\gamma)\rangle}
    +
    \langle P_{10}\rangle_{|\psi(\beta,\gamma)\rangle}
    \bigr),
    \label{eq:experimental_cost_landscape}
\end{equation}
because \(|01\rangle\) and \(|10\rangle\) are the cut states, while
\(|00\rangle\) and \(|11\rangle\) have zero cost. In the ideal \(p=1\) case,
this cost takes the closed form
\begin{equation}
    F(\beta,\gamma)
    =
    -\frac{1}{2}
    + \frac{1}{2}\sin(4\beta)\sin(\gamma),
    \label{eq:closed_form_cost_landscape}
\end{equation}
which implies a \(2\pi\)-periodicity in \(\gamma\) and a \(\pi/2\)-periodicity
in \(\beta\), with ideal range \(F\in[-1,0]\). In particular, lines at
\(\gamma=0,\pi,2\pi,\dots\) and \(\beta=0,\pi/4,\pi/2,\dots\) take the constant
value \(F=-1/2\), while the maximum and minimum occur when
\(\sin(4\beta)\sin(\gamma)=\pm1\).

Fig.~\ref{fig:cost_landscape} compares the measured and simulated landscapes.
The measured data preserves the main low-cost and high-cost regions of the ideal
\(p=1\) structure, so the parameter dependence relevant for variational
optimization is clearly resolved. At the same time, the measured landscape is
less symmetric and has lower contrast than the simulation, indicating the
effects of finite readout noise, calibration imperfections, and control errors.
The average error across the full landscape is \(12.3\%\). The largest
deviations appear predominantly at larger \(\beta\), which is plausible because
the mixing layer is implemented through driven rotations. Larger angles require
stronger control action and therefore increase sensitivity to calibration errors
and decoherence. By contrast, the weaker dependence on \(\gamma\) is consistent
with the corresponding \(z\)-rotation being implemented virtually.

A related qualitative observation was reported in \cite{obst2023comparing},
where a five-qubit MAX-CUT instance was implemented on superconducting and
trapped-ion platforms and compared with simulation for \(p\in\{1,2\}\). That
study likewise found that visible structure and contrast in the measured energy
landscape are important for practical optimization. Our experiment addresses a
much smaller instance on a different platform, but it is consistent with the
same general picture. Even for this minimal two-qubit problem, useful QAOA
behavior relies on retaining a clearly resolvable cost landscape in the measured
data.

\section{Conclusion}

We presented a proof-of-principle implementation of a single-layer QAOA instance
for the smallest nontrivial MAX-CUT problem on a two-qubit NV\(^-\)-center
quantum processor. The experiment combines hardware adaptation of the QAOA
circuit, optical readout modeled by a diagonal observable, and a Hadamard-based
linear-inversion procedure for reconstructing the computational-basis
populations required for cost evaluation.

The measured cost landscape reproduces the main qualitative structure of the
ideal simulation, confirming that the essential ingredients of QAOA can be
realized on this platform, while also exposing its current limitations: finite
readout fidelity, calibration imperfections, untracked phase accumulation, and
decoherence during the two-qubit sequence.

These results establish a baseline protocol for QAOA on room-temperature NV-
center hardware. The most direct next steps are improved phase tracking, more
refined calibration and readout models~\cite{he2024direct}, coherence protection
during long conditional gates \cite{taminiau_universal_2014}, and extension from
this minimal benchmark to larger graph instances and deeper ansätze by coupling
to \({}^{13}\mathrm{C}\) nuclear spins or performing circuit
cutting~\cite{Mitarai_2021}.

\begin{appendices}

\section{Details of the population reconstruction}%
\label{app:population_reconstruction}

Here we derive the linear-inversion formulas used to reconstruct the
computational-basis populations from the fluorescence measurements introduced in
the main text. Starting from the readout model in
Eq.~\eqref{eq:measurement_observable}, with calibration values \(I_s\) from
Eq.~\eqref{eq:calibration_values}, the goal is to recover \(\langle
P_s\rangle_{|\psi\rangle}\) from the four expectation values in
Eq.~\eqref{eq:four_measurements}.

For a single qubit, \(|0\rangle\langle 0|=\frac{1}{2}(I+Z)\) and \(
|1\rangle\langle 1|=\frac{1}{2}(I-Z)\). Therefore, for two qubits,
\begin{equation}
P_s = |s_1s_2\rangle\langle s_1s_2|
= \frac{1}{4} \sum_{t\in\{0,1\}^2} (-1)^{s\cdot t}\, Z_1^{t_1}Z_2^{t_2},
\label{eq:appendix_projector_expansion}
\end{equation}
where \(s\cdot t=s_1t_1+s_2t_2\) is understood modulo \(2\).

Using the spectral decomposition \(O=\sum_s I_s P_s\), the readout observable
can be written as
\begin{equation}
O = \sum_{t\in\{0,1\}^2} c_t\, Z_1^{t_1}Z_2^{t_2},
\label{eq:appendix_O_Z_expansion}
\end{equation}
with coefficients
\begin{equation}
c_t = \frac{1}{4} \sum_{s\in\{0,1\}^2} I_s\,(-1)^{s\cdot t}.
\label{eq:appendix_ct}
\end{equation}
Thus, the coefficients \(c_t\) are given by the Walsh-Hadamard transform of the
calibration values \(\{I_s\}\).

Now let \(|\psi\rangle = U|00\rangle\) and define
\begin{equation}
X^x := X_1^{x_1}X_2^{x_2}, \qquad x\in\{0,1\}^2.
\end{equation}
Using \(XZX=-Z\), the expectation value after a final bit-flip pattern \(X^x\)
becomes
\begin{equation}
\langle O\rangle_{X^x|\psi\rangle}
=
\sum_{t\in\{0,1\}^2}
c_t\,(-1)^{x\cdot t}
\langle Z_1^{t_1}Z_2^{t_2}\rangle_{|\psi\rangle}.
\label{eq:appendix_O_shifted}
\end{equation}
A second Walsh-Hadamard transform then gives
\begin{equation}
\langle Z_1^{t_1}Z_2^{t_2}\rangle_{|\psi\rangle}
=
\frac{1}{4c_t}
\sum_{x\in\{0,1\}^2}
(-1)^{x\cdot t}
\langle O\rangle_{X^x|\psi\rangle},
\qquad t\in\{0,1\}^2,
\label{eq:appendix_Z_reconstruction}
\end{equation}
provided \(c_t\neq 0\).

Substituting these correlators into Eq.~\eqref{eq:appendix_projector_expansion}
yields the reconstructed populations,
\begin{equation}
\langle P_s\rangle_{|\psi\rangle}
=
\frac{1}{4}
\sum_{t\in\{0,1\}^2}
(-1)^{s\cdot t}
\langle Z_1^{t_1}Z_2^{t_2}\rangle_{|\psi\rangle}.
\qquad s\in\{0,1\}^2.
\label{eq:appendix_population_reconstruction}
\end{equation}

\end{appendices}

\section{Acknowledgements}
This work was supported by the German Research Foundation (DFG) as part of
Germany's Excellence Strategy (EXC~2050/2, Project ID~390696704), Cluster of
Excellence ``Centre for Tactile Internet with Human-in-the-Loop'' (CeTI), TUD
Dresden University of Technology; by the Federal Ministry of Research,
Technology, and Space (BMFTR) within the project 6G-life (ID~16KIS2413K) as part
of the research program ``Souverän. Digital. Vernetzt.''; by the BMFTR within
the projects QUARKS (ID~16KIS1998K) and CommUnity (ID~16KISS012K); and by the
project ``Next Generation AI Computing (gAIn)'', funded by the Bavarian Ministry
of Science and the Arts and the Saxon Ministry for Science, Culture, and
Tourism. Large language models (LLMs) were used to assist in drafting and
editing text and code. All generated content was reviewed, verified, and
approved by the authors.


\end{document}